\documentclass[pdflatex,sn-mathphys]{sn-jnl}
\usepackage{graphicx}
\usepackage{hyperref}
\usepackage{pifont}
\usepackage{threeparttable}
\usepackage{makecell}
\usepackage{multicol}
\usepackage{multirow}
\usepackage{booktabs}
\usepackage{makecell}
\usepackage{adjustbox}
\usepackage{bm}

\jyear{2023}%

\theoremstyle{thmstyleone}%
\theoremstyle{thmstyletwo}%

\theoremstyle{thmstylethree}%

\begin{document}

\title[DefiNet]{Modeling crystal defects using defect informed neural networks}

\author[1,2]{\fnm{Ziduo} \sur{Yang}}

\author[2]{\fnm{Xiaoqing} \sur{Liu}}

\author[2]{\fnm{Xiuying} \sur{Zhang}}

\author*[3]{\fnm{Pengru} \sur{Huang}}\email{pengru@nus.edu.sg}

\author[3]{\fnm{Kostya S.} \sur{Novoselov}}

\author*[2,4]{\fnm{Lei} \sur{Shen}}\email{shenlei@nus.edu.sg}

\affil[1]{\orgdiv{Department of Electronic Engineering, College of Information Science and Technology}, \orgname{Jinan University}, \orgaddress{\city{Guangzhou}, \postcode{510632}, \country{China}}}

\affil[2]{\orgdiv{Department of Mechanical Engineering}, \orgname{National University of Singapore}, \orgaddress{\city{9 Engineering Drive 1}, \postcode{117575}, \country{Singapore}}}

\affil[3]{\orgdiv{Institute for Functional Intelligent Materials}, \orgname{National University of Singapore}, \orgaddress{\city{4 Science Drive 2}, \postcode{117544}, \country{Singapore}}}

\affil[4]{\orgdiv{National University of Singapore (Chongqing) Research Institute}, \orgaddress{\city{Chongqing}, \postcode{401123}, \country{China}}}


\abstract{Most AI-for-Materials research to date has focused on ideal crystals, whereas real-world materials inevitably contain defects that play a critical role in modern functional technologies. The defects break geometric symmetry and increase interaction complexity, posing particular challenges for traditional ML models. Here, we introduce Defect-Informed Equivariant Graph Neural Network (DefiNet), a model specifically designed to accurately capture defect-related interactions and geometric configurations in point-defect structures. DefiNet achieves near-DFT-level structural predictions in milliseconds using a single GPU. To validate its accuracy, we perform DFT relaxations using DefiNet-predicted structures as initial configurations and measure the residual ionic steps. For most defect structures, regardless of defect complexity or system size, only 3 ionic steps are required to reach the DFT-level ground state. Finally, comparisons with scanning transmission electron microscopy (STEM) images confirm DefiNet's scalability and extrapolation beyond point defects, positioning it as a \textcolor{black}{valuable} tool for defect-focused materials research.}

\keywords{Defect Calculations, Materials Discovery, Equivariant Graph Neural Networks, Structural Relaxation}



\maketitle

\section{Introduction}
Studying crystalline materials and their devices necessarily requires investigating defects. On the one hand, defects are intrinsic and unavoidable in crystals, often significantly limiting device performance. On the other hand, defect engineering, the deliberate introduction of extrinsic defects into materials, is crucial for unlocking novel properties and functionalities in crystalline materials, enabling advancements in modern functional technologies \cite{davidsson2023absorption, huang2023unveiling, mosquera2023identifying, kazeev2023sparse}. 

\begin{figure}[!h]
  \centering
  \includegraphics[width=11.8cm]{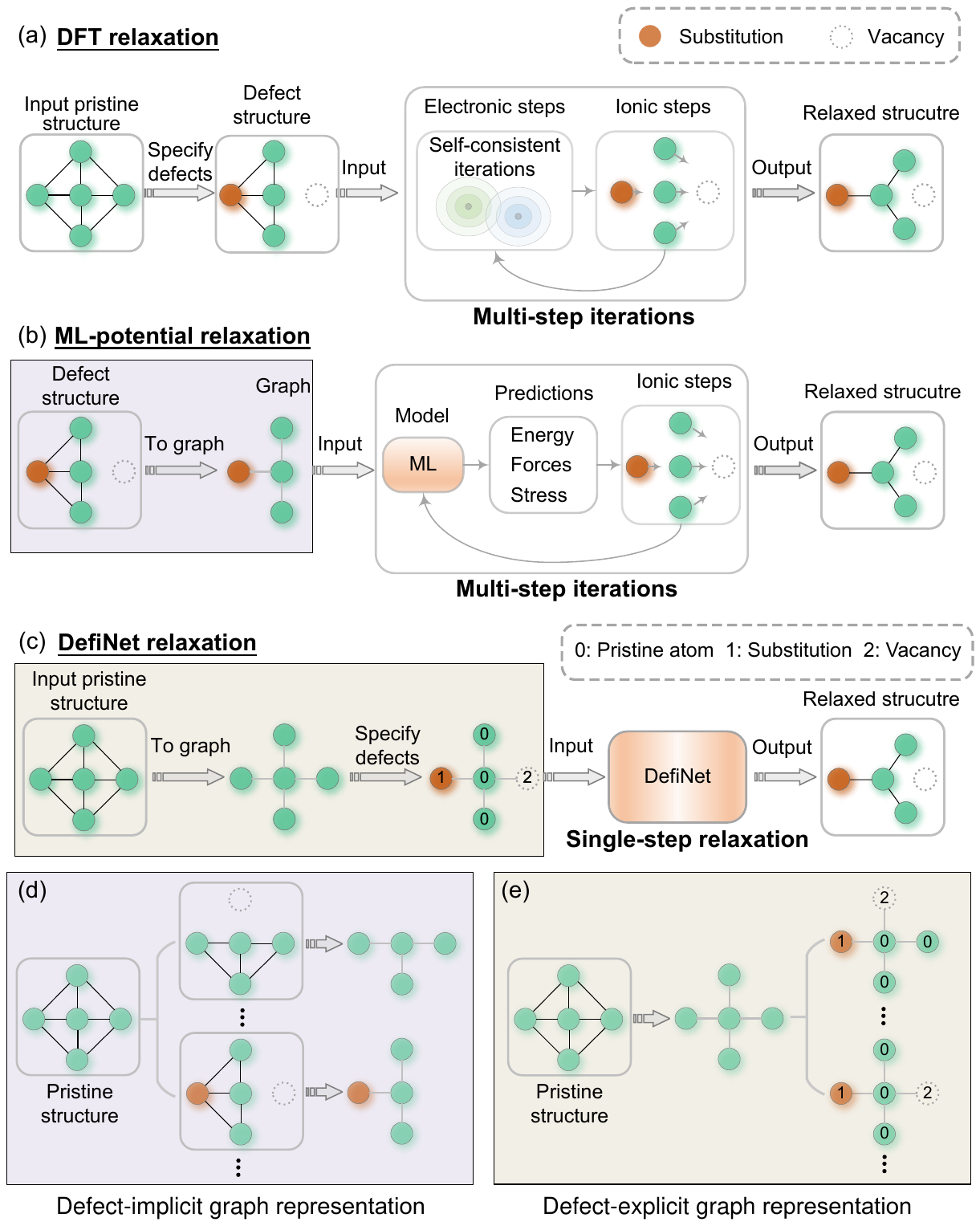}
  \caption{Overview of crystal defect structure relaxation methods. (a) Relaxation using DFT with multi-step iterations. (b) Relaxation using ML potentials with multi-step iterations. (c) Relaxation using our DefiNet with a single step. (d) Defect-implicit graph used by standard GNN workflows, where defect sites are not explicitly labelled. (e) Defect‐explicit graph introduced here, in which nodes carry explicit markers (0 = pristine atom, 1 = substitution, 2 = vacancy) to identify defects.}
  \label{fgr:different_methods}
\end{figure}

The defect space is primarily defined by three variables: the host structure, the types of defects, and defect configurations \cite{huang2023unveiling}. The types of defects are limited to a few categories, such as intrinsic vacancies and impurity substitutions. However, the space for defect configurations is immense, making thorough experimental or computational investigations very challenging \cite{thomas2024substitutional}. These defects typically induce local lattice distortions. To optimize the defect structures, one typically performs conventional \textit{ab initio} methods such as density functional theory (DFT), as depicted in Fig. \ref{fgr:different_methods}(a). DFT calculations involve iterative electronic and ionic steps that gradually converge the system to its lowest energy configuration. These steps are computationally expensive, with the time scaling approximately as $N^3$ where $N$ is the number of atoms, making DFT calculations particularly challenging for large or complex systems.

The emerging technique of machine learning (ML) interatomic potentials \cite{chen2022universal, deng2023chgnet, batzner20223, batatia2022mace, park2024scalable, ko2023recent} has shown the potential in reducing computational demands associated with defect structure optimization. By training a graph neural network (GNN) to iteratively approximate physical quantities such as energies, forces, and stresses, ML-potential relaxation bypasses the computationally intensive electronic step while retaining the ionic step, as shown in Fig. \ref{fgr:different_methods}(b). For example, Mosquera-Lois et al. \cite{mosquera2024machine} and Jiang et al. \cite{jiang2024machine} have demonstrated that ML interatomic potentials can provide both cost-effectiveness and accuracy in identifying the ground-state configurations of defect structures. Despite these advantages, three primary challenges remain in applying ML interatomic potentials to the study of defect structures. First, existing ML interatomic potentials do not explicitly consider the complicated defect-related interactions. Second, the development of ML interatomic potentials heavily relies on the availability of comprehensive databases with detailed labels for energy, forces, or stresses during structural relaxations, which may not always be available for complex defect systems. 

To overcome these challenges, we develop the Defect-Informed Equivariant Graph Neural Network (DefiNet), a single-step ML model specifically designed for the rapid relaxation of defect crystal structures without requiring any iterative process, as shown in Fig. \ref{fgr:different_methods}(c). DefiNet offers four key advantages:

\textcolor{black}{1) Defect-explicit representation---Conventional GNNs model defect structures using defect-implicit graphs, in which no explicit flags denote defect sites and the network must infer them implicitly from structures, as shown in Fig. \ref{fgr:different_methods}(d). DefiNet instead builds a single host-structure graph and attaches markers to nodes to explicitly denote defects, yielding a defect-explicit graph (Fig. \ref{fgr:different_methods}(e)). Combined with our defect-aware message passing scheme, this design captures complex defect–defect and defect–host interactions more accurately.}

2) End-to-end trainability---DefiNet directly maps initial structures to relaxed configurations, enabling efficient end-to-end training and scalable parallel computing capabilities. This makes it highly suitable for large-scale calculations as it completely eliminates iterative relaxation steps.

3) Equivariant representation---The model leverages equivariant representation to ensure that rotational transformations of the input structure are consistently reflected throughout the network's layers and in the final output coordinates, leading to more precise geometric representations.

4) Scalability---It is well known that in conventional DFT or ML interatomic potential approaches, computational cost increases significantly with structural complexity and the total number of atoms due to their reliance on iterative algorithms. In contrast, DefiNet’s single-step and end-to-end design enable it to accurately predict defect structures regardless of defect complexity or system size. 

We evaluated DefiNet on 14,866 defect structures across six widely studied materials, including $\rm MoS_2$, $\rm WSe_2$, h-BN, GaSe, InSe, and black phosphorus (BP), each presenting a variety of defects. Our results show that with just a few hundred training samples per material, DefiNet achieves precise structural relaxation within tens of milliseconds using a single GPU, even without utilizing its parallel computing capabilities. To validate the accuracy and efficiency, we use the original unrelaxed structures and DefiNet-predicted structures as initial configurations for DFT calculations. DefiNet improves the computational efficiency by 87\%, demonstrating DefiNet’s efficiency in identifying energetically favorable configurations. Moreover, DefiNet efficiently scales from small to large systems while maintaining its ability to generalize between high- and low-defect-density scenarios. Comparisons with high-resolution scanning transmission electron microscopy (STEM) images of complex defects, such as line defects, further validate the model’s scalability and extrapolation capabilities beyond point defects. Collectively, these advancements establish DefiNet as a powerful tool for defect-focused materials and device research.

\begin{figure}[!h]
  \centering
  \includegraphics[width=11.8cm]{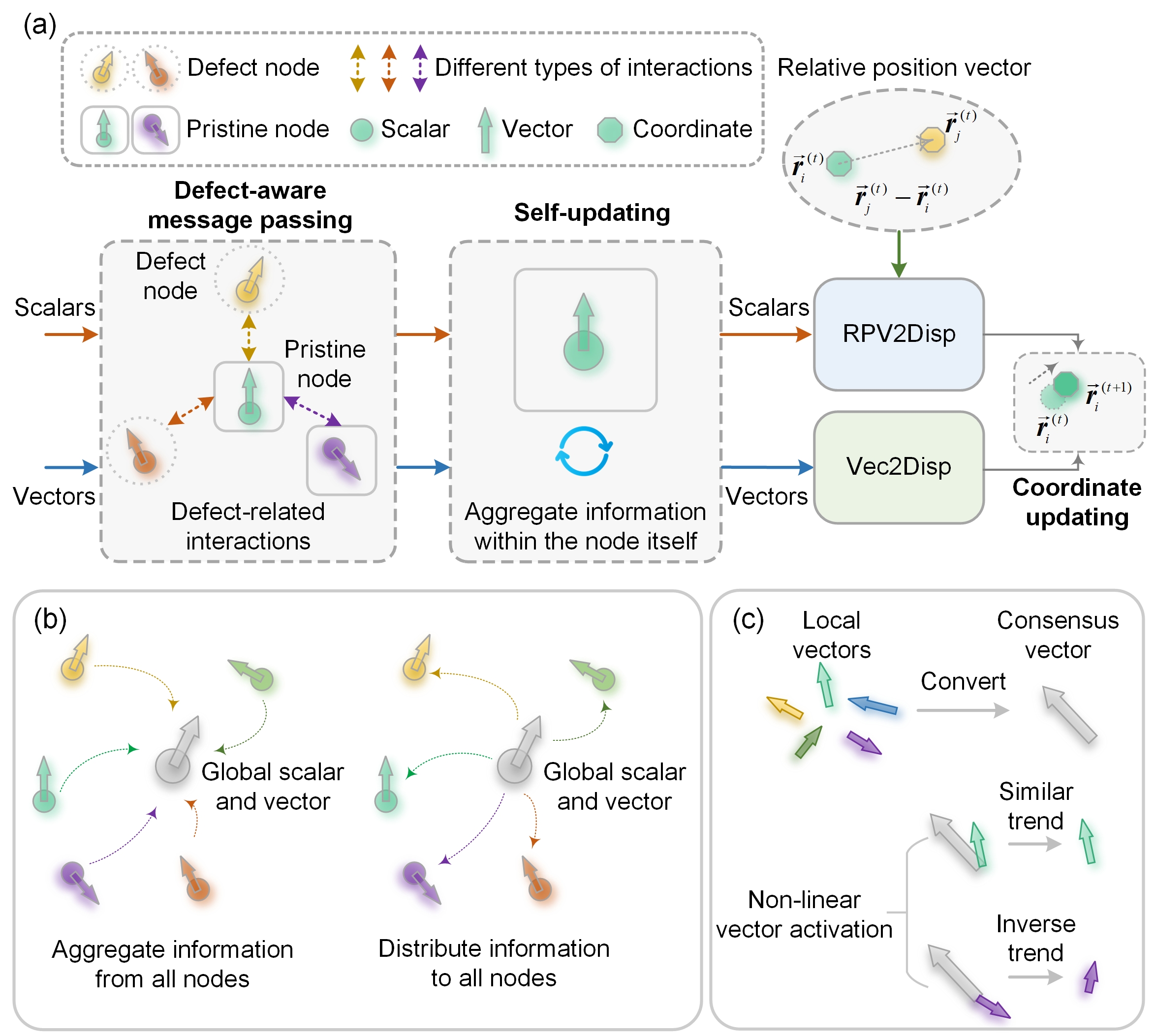}
  \caption{Detailed designs of DefiNet. (a) Overview of the three-stage updating process, including defect-aware message passing, self-updating, and defect-aware coordinate updating. (b) Implementation of global node (including global scalar and global vector). (c) Non-linear vector activation technique.}
  \label{fgr:detailed_design}
\end{figure}

\section{Results}
\subsection{DefiNet architecture}

\textcolor{black}{Graph neural networks (GNNs) operate directly on graph‐structured data, making them ideal for crystalline materials, where atoms map to nodes and interatomic bonds to edges \cite{xie2018crystal, musaelian2023learning, li2022deep, gong2023general, zhong2023transferable, zhong2024universal, zhong2024accelerating, schutt2017schnet, chen2019graph, gasteiger_dimenet_2020, choudhary2021atomistic, chen2022universal, deng2023chgnet, unke2021spookynet, batzner20223, zhang2022interpretable, satorras2021n, omee2022scalable, schutt2021equivariant, han2024survey, dong2025accurate, wu2023graph, li2024local}. DefiNet extends this paradigm with a defect‐explicit representation: instead of relying on defect‐implicit graphs, where defects must be inferred from structures, DefiNet augments a single host‐structure graph with explicit markers (0 = pristine atom, 1 = substitution, 2 = vacancy) to indicate defect sites, thereby enabling the network to explicitly encode defect-related interactions during message passing.} 

The overall architecture of DefiNet is depicted in Fig. \ref{fgr:detailed_design}(a). The model employs a vector-scalar-coordinate triplet representation for each node to encapsulate invariant, equivariant, and structural features, respectively. Scalar features encode information related to the material's properties that are invariant to geometric transformations. Vector features provide geometrical information that is equivariant to rotations. The initial coordinates are updated through successive layers to optimize the structure toward a more stable state.

DefiNet updates this triplet representation through a three-stage graph convolution process, as illustrated in Fig. \ref{fgr:detailed_design}(a). \textcolor{black}{The process begins with defect-aware message passing, in which neighboring nodes exchange information through marker-conditioned edges (i.e., defect–defect, defect–pristine, and pristine–pristine) so that the propagated messages explicitly encode both the presence and the category of each defect.} The self-updating stage then updates the scalar and vector features using the node's internal information. The final stage, defect-aware coordinate updating, optimizes atom coordinates using two specific modules, namely the Relative Position Vector to Displacement (RPV2Disp) and Vector to Displacement (Vec2Disp). These modules predict the necessary displacements to move each atom toward an optimized structure.

DefiNet further incorporates two technologies to boost model performance. First, it adopts the global node (including global scalar and global vector) introduced by Yang et al. \cite{yang2025efficient} to capture long-range interactions, as illustrated in Fig. \ref{fgr:detailed_design}(b). These global components aggregate scalar and vector information from all nodes across the graph and subsequently redistribute it to each node, thereby enhancing the model's ability to identify long-range interactions effectively. Second, while non-linearity is crucial for the expressive power of neural networks, introducing non-linearity into vector representations without compromising equivariance presents a challenge \cite{deng2021vector}. To address this, we have introduced a novel nonlinear vector activation, as illustrated in Fig. \ref{fgr:detailed_design}(c). This method computes a consensus vector by aggregating local vectors, capturing the overarching directional trend among them. Vectors that align with this consensus vector, as indicated by a dot product greater than zero, are deemed significant and retained without changes. In contrast, vectors that diverge from this consensus trend, shown by a dot product less than zero, are modified by adding the consensus vector, thus reorienting them closer to the dominant directional trend. \textcolor{black}{The intuition behind this design is that if most directional features agree on a common trend, then outlier vectors that strongly deviate are likely to be noisy or weakly informative and should be softly regularized toward the consensus.}

\begin{table}[]
\centering
\caption{Overview of point defect types and DFT calculation parameters for the 2DMD datasets}
\label{tbl:dataset}
\begin{tabular}{lllll}
\toprule
Materials   & Substitutions     & Vacancies & Supercell & Cell size ($\rm \AA$)          \\ \midrule
$\rm MoS_2$ & S $\rightarrow$ Se; Mo $\rightarrow$ W        & Mo; S     & $8 \times 8$     & (25.52, 25.52, 20) \\
$\rm WSe_2$ & Se $\rightarrow$ S; W $\rightarrow$ Mo        & W; Se     & $8 \times 8$     & (26.62, 26.62, 20) \\
h-BN        & B $\rightarrow$ C; N $\rightarrow$ C          & B; N      & $8 \times 8$     & (20.10, 20.10, 20) \\
GaSe        & Ga $\rightarrow$ In; Se $\rightarrow$ S       & Ga; Se    & $6 \times 6$     & (22.91, 22.91, 20) \\
InSe        & In $\rightarrow$ Ga; Se $\rightarrow$ S       & In; Se    & $6 \times 6$     & (24.58, 24.58, 20) \\
BP          & P $\rightarrow$ N & P         & $6 \times 6$     & (19.80, 27.61, 20) \\ \bottomrule
\end{tabular}
\end{table}

\subsection{Database}
We have developed a database for 2D material defects (2DMD) \cite{ huang2023unveiling, kazeev2023sparse}, to facilitate the training and evaluation of ML models for defect structure analysis. This database includes structures with point defects for commonly used 2D materials including $\rm MoS_2$, $\rm WSe_2$, h-BN, GaSe, InSe, and black phosphorous (BP). Details of these point defects with supercell specifications are presented in Table \ref{tbl:dataset}. \textcolor{black}{All defects in our dataset are in the neutral charge state.}

\textcolor{black}{The database is divided into two sections: one with a low-density of structured defect configurations, and another with a high-density of randomly configured defects, according to the defect concentration. The low-density section includes 5,933 structures each for $\rm MoS_2$ and $\rm WSe_2$, with defect concentration lower than 1.6\% (1 to 3 defects) per structure, covering all potential configurations within an $8 \times 8$ supercell. The high-density section comprises randomly generated substitution and vacancy defects across all six materials. For each defect concentration—2.5\%, 5\%, 7.5\%, 10\%, and 12.5\%—100 structures were created, resulting in a total of 500 configurations per material and 3000 in total.} In total, the dataset contains 14,866 structures, each comprising 120–192 atoms after applying supercell expansion.

\textcolor{black}{The database is stratified by material and defect density (low vs. high) and then randomly split into training, validation, and test sets in an 8:1:1 ratio. Each subset maintains the same overall data distribution but contains non-overlapping defect configurations.}

\subsection{Evaluation Metric}
We use the coordinate MAE between the ML-relaxed and DFT-relaxed structures to evaluate the model's performance. Since structural variations between unrelaxed and relaxed defect structures are primarily localized near the defect sites, we further introduce localized MAE statistics for a more precise assessment of model's performance. Specifically, we denote atoms within an $x \ \rm{\AA}$ radius of the defect sites as A$_x$, where $x$ is set to 3, 4, 5, and 6 in our experiments. For example, the coordinate MAE for A$_5$ considers only atoms within a 5 $\rm{\AA}$ radius of the defect site when calculating the MAE.

\subsection{Model performance on structures with low-density defects}
We first benchmark DefiNet on structures with low-density defects (defect concentration below 1.6\%), comparing it against the state-of-the-art (SOTA) single-step ML model, DeepRelax \cite{yang2024scalable}. \textcolor{black}{A concise comparison of the key differences between DefiNet and DeepRelax is provided in Supplementary Note 2.} As a baseline, we introduce a Dummy model that simply returns the input initial structure as its output, serving as a control reference for evaluation. All models are trained, validated, and tested on identical datasets.

\begin{figure}[!p]
  \centering
  \includegraphics[width=11.8cm]{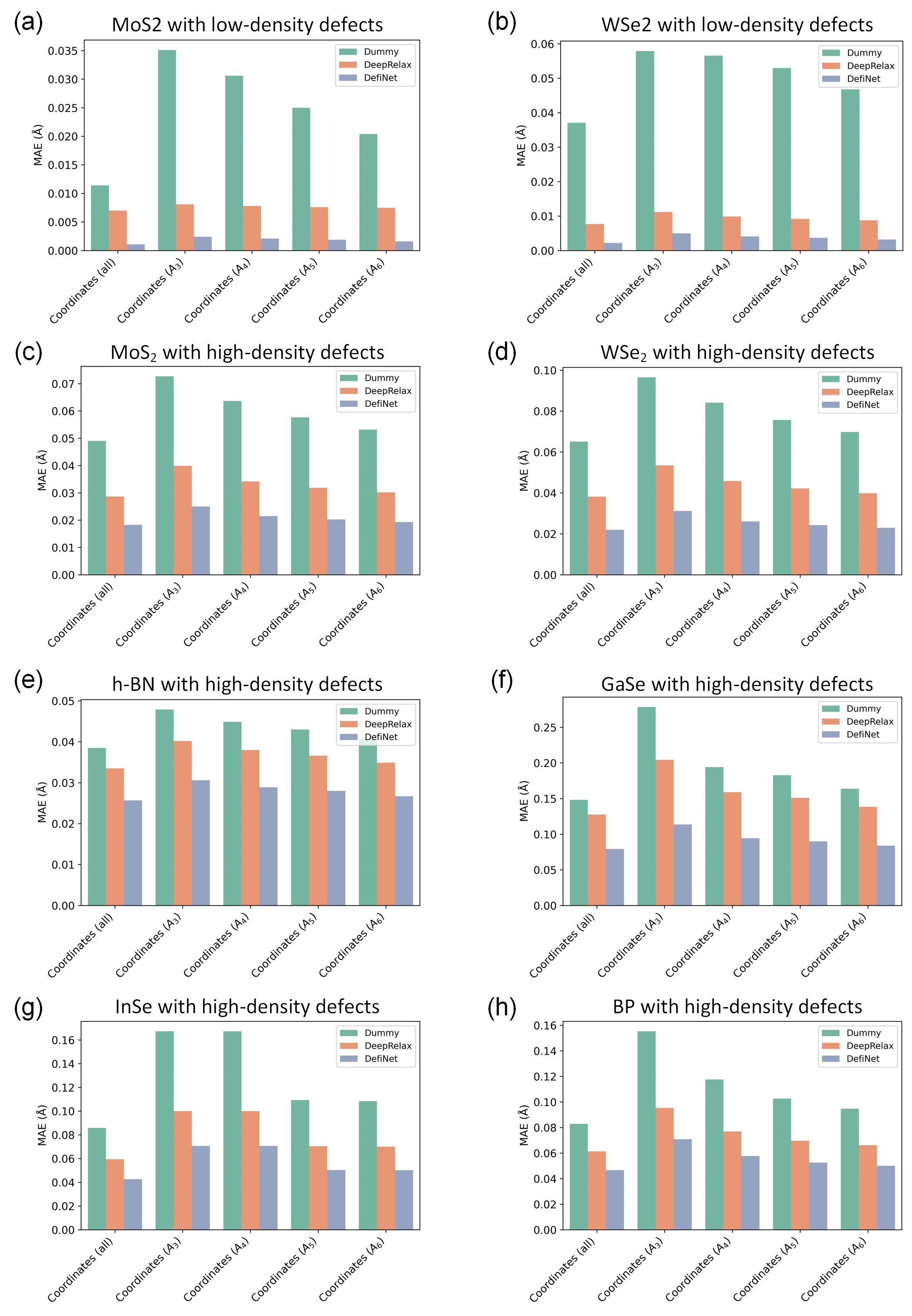}
  \caption{Model performance for structures with low- and high-density defects. (a) $\rm MoS_2$ and (b) $\rm WSe_2$ with low-density defects, and (c) $\rm MoS_2$, (d) $\rm WSe_2$, (e) h-BN, (f) GaSe, (g) InSe, and (h) BP with high-density defects. A$_3$, A$_4$, A$_5$, and A$_6$ represent MAE calculations using only atoms within $3 \rm{\AA}$, $4 \rm{\AA}$, $5 \rm{\AA}$, and $6 \rm{\AA}$ radii around defect sites, respectively. }
  \label{fgr:all_performance}
\end{figure}

Fig. \ref{fgr:all_performance}(a)-(b) presents the performance of the models, showing that both DeepRelax and DefiNet significantly outperform the Dummy model. DefiNet surpasses DeepRelax notably, achieving improvements of 78.38\%, 61.86\%, 64.77\%, 66.67\%, and 70.37\% in coordinate MAE for all atoms, A$_3$, A$_4$, A$_5$, and A$_6$, respectively, across all defect structures in both materials. Additionally, DefiNet is approximately 26.2 times more computationally efficient than DeepRelax in terms of inference speed, as shown in Supplementary Table 1. 

\textcolor{black}{We also assess DefiNet's performance using different percentages of the training data, as shown in Supplementary Fig. 9, to investigate the relationship between dataset size and model accuracy. The results show that performance improves rapidly when increasing the training size in the low-data regime (e.g., from 10\% to 30\%), but the gains become increasingly marginal beyond that point. This trend suggests that DefiNet can learn effectively from limited data, while additional data primarily serves to fine-tune predictions rather than drive major improvements.} 

\subsection{Model performance on structures with high-density defects}
While low-density defects are more commonly studied, they represent only a small portion of the entire defect space. High-density defects can reveal important and unique physical phenomena that low-density studies may not capture. In particular, interactions between multiple defects can significantly influence material properties in ways that isolated defects cannot. These complex defect-related interactions pose a significant challenge for ML models.

\textcolor{black}{Here, we demonstrate that DefiNet also achieves strong performance on structures with high-density defects (defect concentrations between 2.5\% and 12.5\%), as shown in Fig. \ref{fgr:all_performance}(c)-(h).} We make three key observations: First, DefiNet proves to be robust across multiple materials. Second, compared to the results in Fig. \ref{fgr:all_performance}(a)-(b), both DeepRelax and DefiNet show less significant improvements. This is likely due to two factors: (1) the high-density defect datasets contain significantly fewer samples (only 500 per material), limiting learning capacity; and (2) the space of possible defect configurations increases substantially with defect density, making the task more complex. Third, DefiNet still significantly outperforms DeepRelax, with improvements of 32.82\%, 35.88\%, 34.08\%, 33.88\%, and 33.33\% in coordinate MAE for all atoms, A$_3$, A$_4$, A$_5$, and A$_6$ respectively, across all defect structures in the six materials.

\begin{figure}[!ht]
  \centering
  \includegraphics[width=11.8cm]{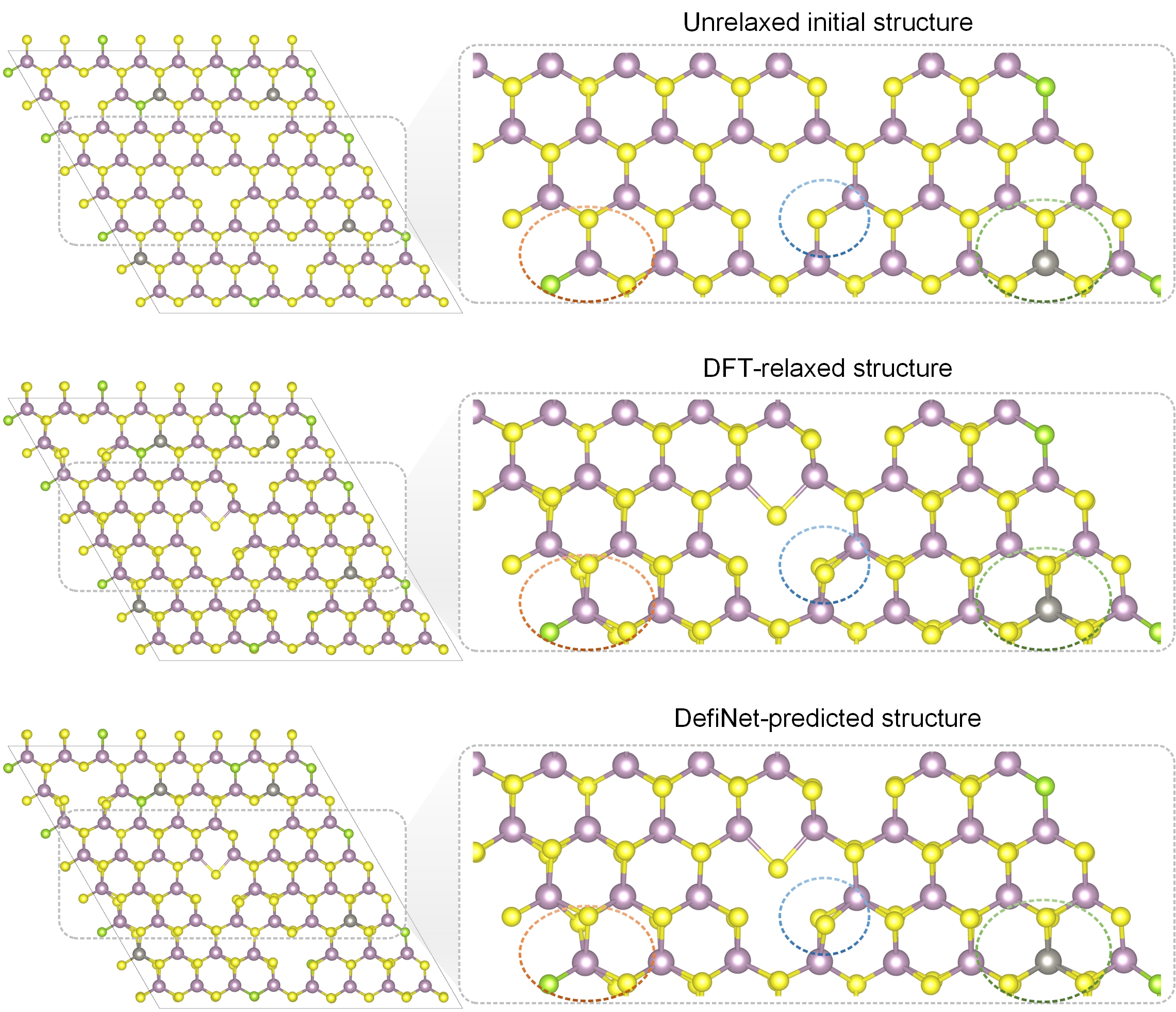}
  \caption{Example of an $\rm MoS_2$ crystal structure containing both substitutional and vacancy defects, alongside the corresponding DFT-relaxed and DefiNet-predicted structures. }
  \label{fgr:vis_example}
\end{figure}

Fig. \ref{fgr:vis_example} provides a visual comparison of the unrelaxed, DFT-relaxed, and DefiNet-predicted structures. As can be seen, the DefiNet-predicted structure closely matches the DFT-relaxed structure, demonstrating the model's effectiveness in handling complex defect configurations. 

\begin{figure}[!h]
  \centering
  \includegraphics[width=11.8cm]{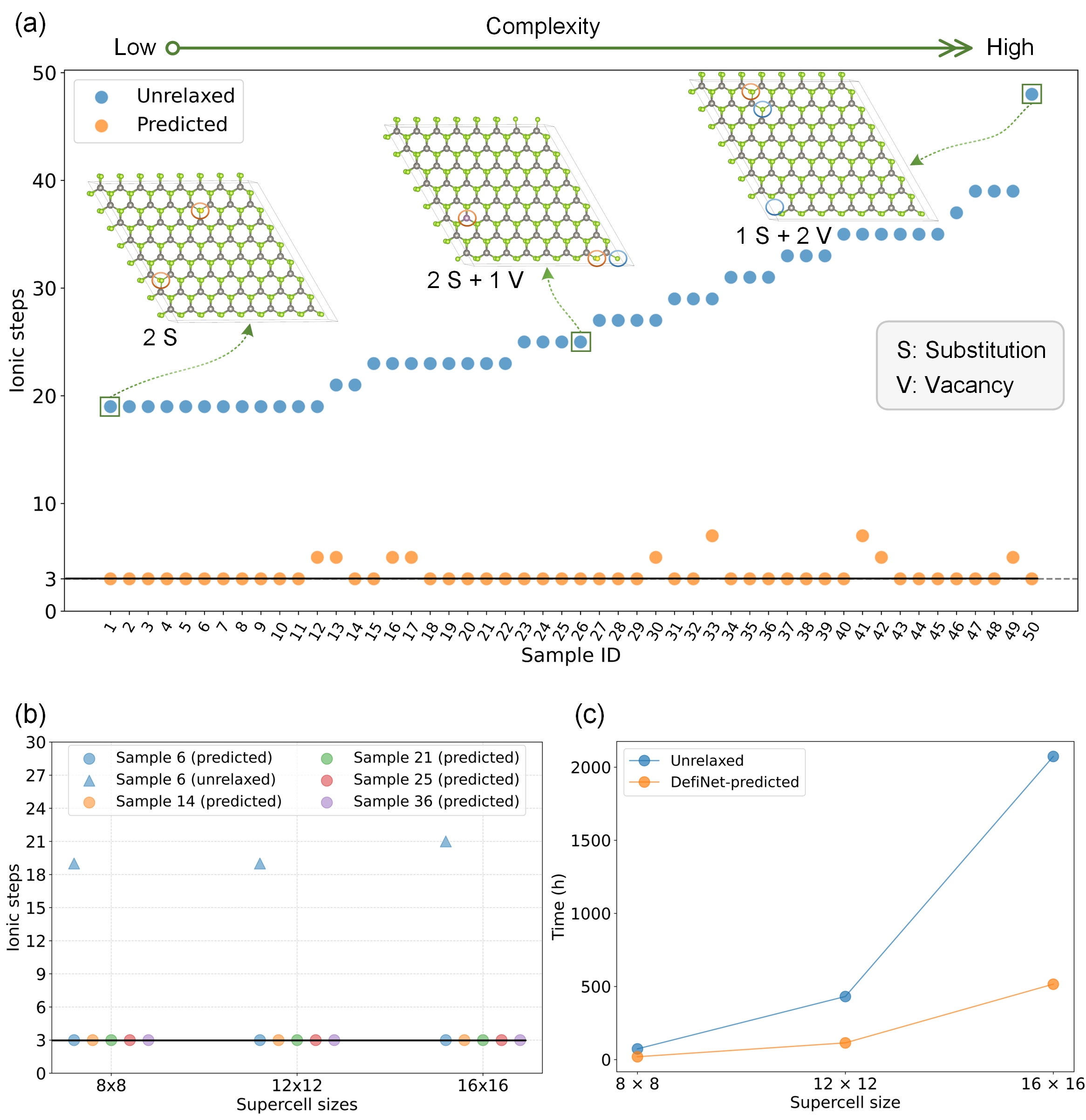}
  \caption{DFT validation on DefiNet's accuracy, efficiency, and scalability. (a) Comparison of the number of DFT ionic steps required to relax structures starting from the initial unrelaxed configurations and from the DefiNet-predicted structures for low-density defects. The steady residual ionic steps against the defect complexity are indicated by a horizontal black solid line. The sample ID is sorted based on the number of ionic steps required by the unrelaxed structures for better observation. (b) Residual ionic steps for five randomly selected defect structures from the 50 samples across different supercell sizes, starting from DefiNet-predicted configurations. \textcolor{black}{Only a single reference run is shown for unrelaxed structures due to the high computational cost of initiating DFT relaxation from unrelaxed configurations.} The steady residual ionic steps against the structural size are indicated by a horizontal black solid line. (c) Comparison of DFT CPU core hours on large supercells using unrelaxed and DefiNet-predicted configurations. Due to the extremely high computational cost associated with the unrelaxed structure of the $16 \times 16$ supercell size with 770 atoms, only one sample was selected as an example for this experiment.}
  \label{fgr:dft}
\end{figure}

\subsection{DFT validation}
Validating the energetic favorability of ML-predicted structures is essential to ensuring their physical relevance, accuracy, and efficiency. While coordinate errors provide insight into geometrical accuracy, further analysis is needed to confirm that the predicted structures correspond to local minima on the potential energy surface. We conduct DFT validations to assess whether the structures relaxed by DefiNet are the same as or very similar to DFT ones. \textcolor{black}{For this validation, we randomly selected 25 WSe$_2$ and 25 MoS$_2$ structures from the low-density defect test set for DFT calculations. Detailed settings for the DFT calculations are provided in Section \ref{sec:dft_calc}. These two materials were chosen because they appear in both low- and high-density defect categories, making them well-suited for evaluating DFT validation across different defect densities.}

We compared the number of ionic steps required for convergence in two cases: starting from unrelaxed structures and starting from DefiNet-predicted structures. The results, shown in Fig. \ref{fgr:dft}(a), indicate that using DefiNet-predicted structures as starting points significantly reduces the computational effort required for DFT relaxation, with the number of ionic steps decreasing by approximately 87\%. Notably, these residual ionic steps also remain nearly constant, regardless of defect complexity. The very low residual ionic steps demonstrate the high accuracy of DefiNet. The steady residual ionic steps, even for highly complex defects, highlight the exceptional efficiency of DefiNet. \textcolor{black}{Importantly, both initialization strategies (starting from unrelaxed structures and from DefiNet-predicted configurations) converge to the same final DFT-relaxed configurations, with a coordinate MAE of zero across all samples.} Additional DFT validation results for high-density defect scenarios are available in Supplementary Fig. 1, which also demonstrates DefiNet’s promising performance. 

To evaluate the scalability of DefiNet, we tested its performance across different supercell sizes. Specifically, we randomly selected five defect structures from the test set containing different types of defects. We then created supercells with sizes of $8 \times 8$, $12 \times 12$, and $16 \times 16$, resulting in structures with around 190 atoms, 430 atoms, and 770 atoms, respectively. DefiNet was used to predict the relaxed structures for these unrelaxed configurations. We assessed both the residual ionic steps and the CPU core hours required for the DefiNet-predicted structures, comparing these results to those of the unrelaxed structures. 

As illustrated in Fig. \ref{fgr:dft}(b), DefiNet consistently achieves constant ionic steps of 3, irrespective of the system size, demonstrating its ability to scale effectively with increasing system size. We further compare the CPU core hours required for the relaxation of both unrelaxed and DefiNet-predicted structures. As shown in Fig. \ref{fgr:dft}(c), the computational cost for the large-scale unrelaxed structure is extremely high. In contrast, the relaxation time for the DefiNet-predicted structures is significantly reduced, highlighting DefiNet's capability for large systems by dramatically decreasing the computational cost. Further scalability evaluations are detailed in Supplementary Note 5.

\subsection{Experimental validation}

\begin{figure}[!h]
  \centering
  \includegraphics[width=11.8cm]{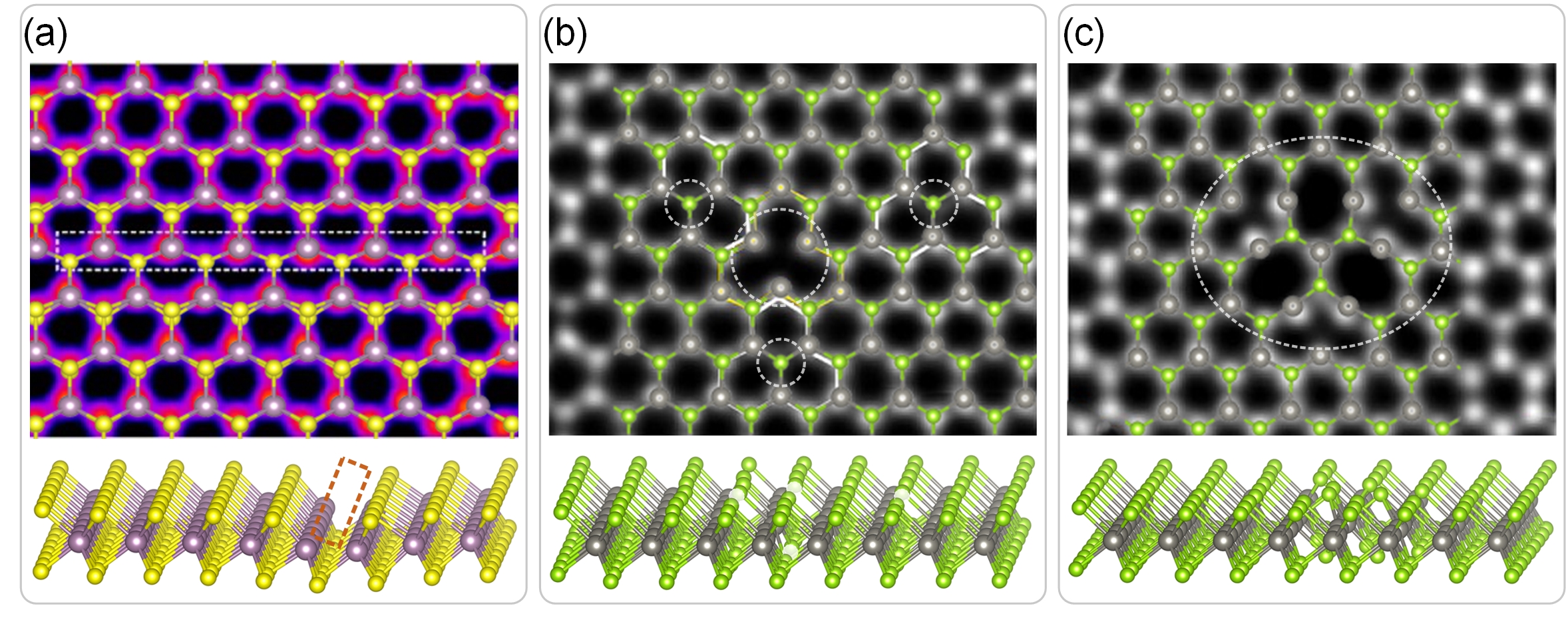}
  \caption{Comparison between STEM images and DefiNet-relaxed structures. \textcolor{black}{Ball-and-stick models of the corresponding DefiNet-relaxed structures are shown below each STEM image, with a line defect marked by orange rectangles.} (a) STEM image of MoS$_2$ featuring a line defect (sequential S vacancies), overlaid with the DefiNet-relaxed structure. Reprinted with permission from \cite{wang2016detailed}. Copyright 2016 American Chemical Society. (b) STEM image of WSe$_2$ with mixed $\rm SV_{Se}$ and $\rm DV_{Se}$ defects, overlaid with the DefiNet-relaxed structure. (c) STEM image of WSe$_2$ with a three-fold symmetrical trefoil defect, overlaid with the DefiNet-relaxed structure. Defect sites are highlighted with white dotted lines for clarity.}
  \label{fgr:stem}
\end{figure}

To further validate the accuracy and extrapolation of DefiNet using experimental results, we conducted comparisons with STEM images, assessing the alignment between DefiNet-relaxed structures and actual experimental observations. Fig. \ref{fgr:stem}(a)-(c) shows STEM images (overlaid with the DefiNet-relaxed structure) of MoS$_2$ \cite{wang2016detailed} and WSe$2$ \cite{lin2015three} with different types of complex defects, including in a line defect (sequential S vacancies), mixed single Se ($\rm SV_{Se}$) with double Se vacancies ($\rm DV_{Se}$), and a three-fold symmetric trefoil defect. The strong alignment between the DefiNet-predicted and experimentally observed structures highlights DefiNet’s accuracy and extrapolation in capturing such complex defects beyond the training point defects. \textcolor{black}{We provide a comparison among the unrelaxed structures, DefiNet-predicted structures, and the STEM image, as shown in Supplementary Fig. 4.}

\subsection{Comparison to ML-potential relaxation}
ML-potential relaxation is a popular alternative to DFT-based relaxation methods. To demonstrate the superiority of DefiNet, we compare it against two well-known ML-potential models: M3GNet and CHGNet. These methods typically require large datasets to train GNN surrogate models that iteratively approximate physical quantities such as energies, forces, and stresses. For this comparison, we used the $\rm MoS_2$ low-density defect dataset, which contains a sufficient number of samples (5,933) with detailed information obtained during DFT-based relaxation. All methods were trained, validated, and tested on the same data splits. Detailed experimental settings are provided in Supplementary Note 7. As shown in Supplementary Fig. 5, DefiNet significantly outperforms M3GNet, CHGNet, and DeepRelax in terms of coordinate MAE and robustness. This result is further validated by DFT calculations, with detailed comparisons available in Supplementary Fig. 6.

\subsection{Ablation study}
To elucidate the contributions of DefiNet's key architectural components, we performed an ablation study focusing on its two main innovations:
\begin{itemize}
    \item Defect-Aware Message Passing (DAMP): This component allows the model to capture complex interactions involving defects.
    \item Defect-Aware Coordinate Updating (DACU): The RPV2Disp and Vec2Disp modules are designed to update atomic coordinates effectively, taking into account the unique influences of defects on the surrounding lattice.
\end{itemize}
We created two ablated versions of DefiNet to assess the impact of these components:
\begin{itemize} 
    \item Vanilla Model: This version removes both main components, DAMP and DACU. 
    \item Vanilla + DAMP: This version includes the DAMP but removes the Defect-Aware Coordinate Updating modules. 
    \item Vanilla + DAMP + DACU (DefiNet): This is the full DefiNet model incorporating both components. 
\end{itemize}

The results on high-density datasets, as shown in Supplementary Fig. 7, indicate that both ablated models exhibit decreased performance compared to the full DefiNet. These findings confirm that both components are critical for DefiNet's superior performance.

\textcolor{black}{We also conduct an additional ablation study to evaluate two auxiliary components: global nodes and nonlinear vector activation. As shown in Supplementary Fig. 8, both mechanisms improve model performance, supporting their inclusion in the final architecture.}

\section{Discussion}
Recently, GNNs have been used for defect property and structure analysis \cite{mosquera2024machine, jiang2024machine, witman2023defect, way2024defect, mannodi2022universal, frey2020machine, rahman2024accelerating, fang2024leveraging}, showing great potential to reduce the high computational cost of DFT calculations. Two recent works \cite{mosquera2024machine, jiang2024machine} have demonstrated that employing machine learning (ML) interatomic potentials can achieve both cost-effectiveness and accuracy in searching ground-state configurations of defect structures. \textcolor{black}{Those approaches, however, require large databases annotated with energies, forces, and stresses, and they treat defect sites only implicitly, leaving the network to infer defect–defect interactions on its own. DefiNet avoids these limitations. First, it is trained solely on pairs of initial and relaxed structures, which makes it easier to implement in real applications. Second, it explicitly considers complex defect-related interactions, leading to more accurate relaxation of defect crystal structures. We also benchmark DefiNet against the previous single-step model, DeepRelax. DefiNet not only achieves a significantly lower coordinate MAE but also runs nearly 26$\times$ faster than DeepRelax.} 

Our scalability tests demonstrate that DefiNet maintains high accuracy when applied to larger systems beyond the sizes used during training. \textcolor{black}{Moreover, we perform two transferability evaluations: (1) Train DefiNet on high-density defect structures and test on low defect-density structures, and vice versa. (2) Train DefiNet on structures with short average defect–defect distances and test on those with long distances, and vice versa. These experiments demonstrate DefiNet’s good transferability (see Supplementary Note 9).}

The DFT validations confirm that the structures predicted by DefiNet are energetically favorable. Importantly, initiating DFT calculations from DefiNet-predicted structures significantly reduces the number of required ionic steps by approximately 87\%, irrespective of defect complexity or system size. This hybrid approach leverages the speed of DefiNet and the precision of DFT, offering an efficient pathway for exploring defect structures in materials. While DefiNet demonstrates remarkable performance, certain limitations warrant discussion.

\textcolor{black}{First, this study focuses exclusively on 2D materials with point defects, and only six materials comprising a limited subset of elements from the periodic table are considered. As a result, the trained DefiNet model cannot be directly generalized to materials containing previously unseen elements. Expanding DefiNet to support a broader range of materials, including both 2D and 3D systems, as well as more complex defect types, would significantly enhance its applicability and generalization capability.}

\textcolor{black}{Second, in this work, we only focus on defects in the neutral charge state. It is worth noting that point defects in semiconductors frequently adopt multiple charge states, each with distinct geometric relaxations. Because existing ML approaches struggle to encode charge directly, most studies to date also limit themselves to neutral or fixed ionic states \cite{kavanagh2024identifying, jiang2024machine}. There are two possible directions for extending DefiNet to charged defects: (1) transfer learning from a neutral‐trained model to charged configurations, or (2) introducing a global charge‐state embedding as an additional input feature. Unfortunately, the lack of sufficiently large, labeled datasets of charged‐defect geometries prevents us from exploring these strategies here, and we therefore leave this as an important direction for future work.}

\textcolor{black}{Third, point defects in low-symmetry semiconductors can occupy several energetically competitive local minima (i.e., metastable configurations) with distinct geometries and functional behaviors \cite{huang2023metastability,kavanagh2022impact}. Since DefiNet outputs only a single relaxed structure per defect, the current version of DefiNet is unable to capture these alternative metastable states.}

\section{Methods}

\subsection{Input representation}
In this work, the defect structure is represented as a defect-explicit graph $\mathcal{G} = (\mathcal{V}, \mathcal{E}, \mathcal{M})$, where $\mathcal{V}$ and $\mathcal{E}$ are sets of nodes and edges corresponding to atoms and bonds within the pristine structure, and $\mathcal{M}$ is a set of markers representing defect types. Each marker $m_i \in \mathcal{M}$ is a categorical variable that takes a value from the set $\{0, 1, 2\}$, where $0$ denotes a pristine atom, $1$ indicates a substitution, and $2$ represents a vacancy. \textcolor{black}{By contrast, a conventional defect-implicit graph \(\mathcal{\tilde{G}} = (\mathcal{V}, \mathcal{E})\) omits defect information \(\mathcal{M}\). In principle, a sufficiently expressive GNN could infer defect sites from structures alone, but doing so is often inefficient, as representation learning is empirically data-hungry \cite{shalev2014understanding}. Providing explicit markers imposes a strong inductive bias: the network no longer has to learn a feature extractor that separates pristine atoms from defect sites, enabling the model to reach the same generalization error with fewer training examples. Importantly, for vacancies, a placeholder node is retained at the position of the missing atom in the pristine lattice and marked with $m_i = 2$. This node is treated as an active part of the graph and participates in message passing. By explicitly incorporating vacancy sites into the graph structure, the model can directly learn spatial relationships between vacancies and neighboring atoms, rather than relying on implicit inference from the structure.
}

Each node $v_i \in \mathcal{V}$ contains three feature types: scalar $\bm{x}_i \in \mathbb{R}^F$, vector $\vec{\bm{x}}_i \in \mathbb{R}^{F \times 3}$, and coordinates $\vec{\bm{r}}_i \in \mathbb{R}^3$, which encapsulate invariant, equivariant, and structural features, respectively. The number of features $F$ is kept constant throughout the network. The scalar feature is initialized as an embedding dependent solely on the atomic number, given by $\bm{x}_i^{(0)}=E(z_i) \in \mathbb{R}^F$, where $z_i$ is the atomic number and $E$ is an embedding layer that takes $z_i$ as input and returns an $F$-dimensional feature. The vector feature is initially set to $\vec{\bm{x}}_i^{(0)}=\vec{\bm{0}} \in \mathbb{R}^{F \times 3}$. To capture long-range interactions, we introduce a global node \(v_\mathcal{G}\), which includes a global scalar \(\bm{x}_\mathcal{G} \in \mathbb{R}^F\) and a global vector \(\vec{\bm{x}}_\mathcal{G} \in \mathbb{R}^{F \times 3}\). These are initialized as a trainable \(F\)-dimensional feature and \(\vec{\bm{0}}\), respectively. We also define the relative position vector as $\vec{\bm{r}}_{ij}=\vec{\bm{r}}_j-\vec{\bm{r}}_i$ to introduce directional information into the edges. Each node is connected to its closest neighbors within a cutoff distance $D$, with a maximum number of neighbors $N$, where $D$ and $N$ are predefined constants.

\subsection{DefiNet workflow}
\begin{figure}[!h]
  \centering
  \includegraphics[width=11.8cm]{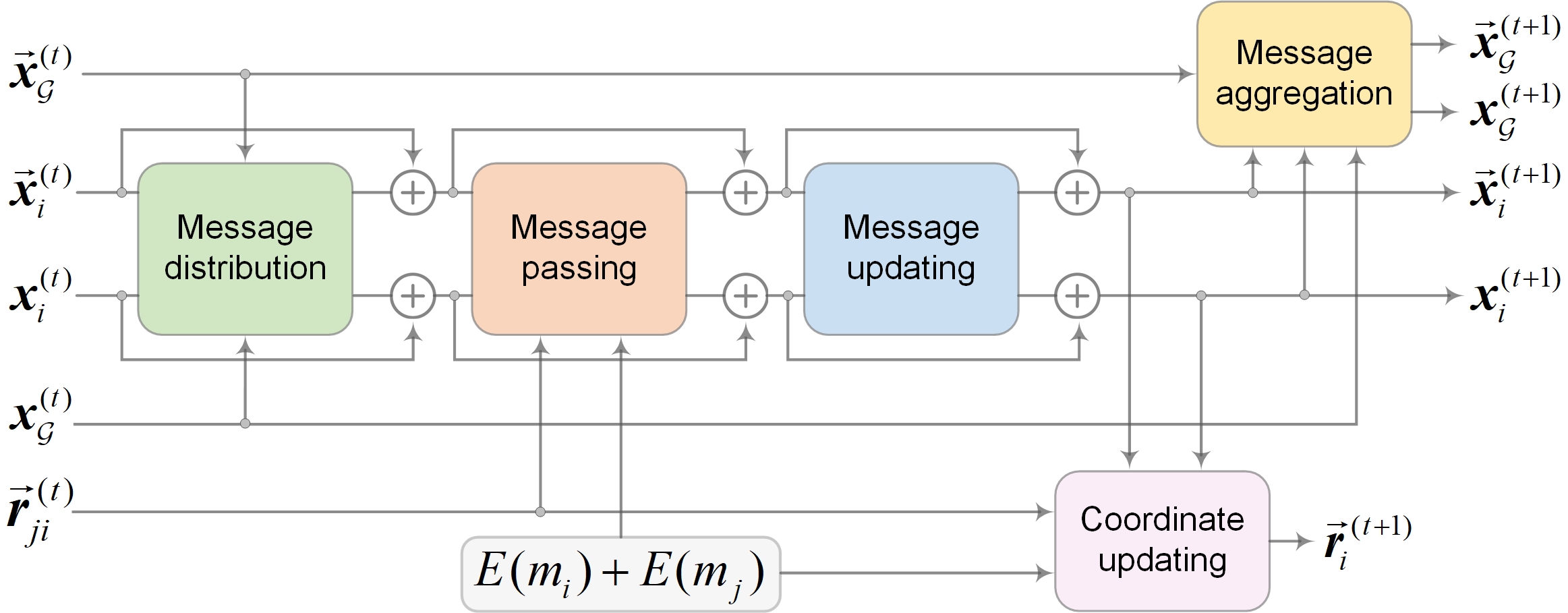}
  \caption{Workflow of the $t$-th graph convolution layer in DefiNet. The process begins with message distribution, where the global scalar \(\bm{x}_{\mathcal{G}}^{(t)}\) and global vector \(\vec{\bm{x}}_{\mathcal{G}}^{(t)}\) are globally distributed to each scalar \(\bm{x}_i^{(t)}\) and vector \(\vec{\bm{x}}_i^{(t)}\). This is followed by defect-aware message passing, which locally collects messages from neighboring nodes \(v_j\), weighting messages according to interatomic distances and the defect markers $m_i$ and $m_j$. Next, message updating refines the node representation using the information within the node itself, resulting in \(\bm{x}_i^{(t+1)}\) and \(\vec{\bm{x}}_i^{(t+1)}\). Coordinate updating then further refines the atomic coordinates, resulting in the updated coordinates \(\vec{\bm{r}}_i^{(t+1)}\). Finally, message aggregation is performed to update the global scalar and vector, resulting in \(\bm{x}_{\mathcal{G}}^{(t+1)}\) and \(\vec{\bm{x}}_{\mathcal{G}}^{(t+1)}\).}
  \label{fgr:DefiNet}
\end{figure}

The proposed DefiNet consists of four layers, each of which updates the node representation through a three-stage graph convolution process that includes defect-aware message passing, self-updating, and defect-aware coordinate updating. This process incorporates message distribution and aggregation to capture long-range interactions, as illustrated in Fig. \ref{fgr:DefiNet}. 

\subsection{Defect-aware message passing}

At layer \(t\) each node \(v_i\) aggregates information from its neighbours \(v_j\)
in a defect-aware manner. This process results in intermediate scalar and vector variables \(\bm{q}_i\) and \(\vec{\bm{q}}_i\), defined as follows:
\begin{equation}
    \bm{q}_i= \sum_{v_j \in \mathcal{N}(v_i)} \phi_h (\bm{x}_{j}^{(t)}) \circ \lambda_h(\Vert \vec{\bm{r}}_{ji}^{(t)} \Vert) \circ \gamma_h \big(E(m_i) + E(m_j)\big) \label{eqn:s_update}
\end{equation}
\begin{align}
    \vec{\bm{q}}_i= & \sum_{v_j \in \mathcal{N}(v_i)} \phi_u( \bm{x}_{j}^{(t)}) \circ \lambda_u(\Vert \vec{\bm{r}}_{ji}^{(t)} \Vert) \circ \gamma_u \big(E(m_i) + E(m_j)\big) \circ \vec{\bm{x}}_j^{(t)} \nonumber \\
    & + \phi_v (\bm{x}_{j}^{(t)}) \circ \lambda_v(\Vert \vec{\bm{r}}_{ji}^{(t)} \Vert) \circ \gamma_v \big(E(m_i) + E(m_j)\big) \circ \frac{\vec{\bm{r}}_{ji}^{(t)}}{\Vert \vec{\bm{r}}_{ji}^{(t)} \Vert}\label{eqn:vec_update}
\end{align}
\textcolor{black}{Here, $\circ$ denotes the element-wise product}, $E$ is an embedding layer that maps the marker $m_i$ to an $F$-dimensional feature, and $\phi_h$, $\phi_u$, $\phi_v$, $\gamma_h$, $\gamma_u$, and $\gamma_v$ are multilayer perceptrons (MLPs). The functions $\lambda_h$, $\lambda_u$, and $\lambda_v$ are linear combinations of Gaussian radial basis functions \cite{schutt2017schnet}. \textcolor{black}{The pair-wise gate \(\gamma(\cdot)\) re-weights each message according to the marker pair \((m_i, m_j)\), thereby distinguishing pristine–pristine, defect–pristine, and defect–defect interactions.}

\subsection{Self-updating}
We employ the self-updating mechanism proposed by Yang et al. \cite{yang2025efficient}. During this phase, the $F$ scalars and $F$ vectors within \(\bm{q}_i\) and \(\vec{\bm{q}}_i\), respectively, are aggregated to generate the updated scalar \(\bm{x}_i^{(t+1)}\) and vector \(\vec{\bm{x}}_i^{(t+1)}\). Specifically, the scalar representation \(\bm{x}_i^{(t+1)}\) and vector representation \(\vec{\bm{x}}_i^{(t+1)}\) are updated according to the following equations:

\begin{equation}
    \bm{x}_i^{(t+1)} = \phi_{s}\big(\bm{q}_{i} \oplus \Vert \bm{V}  \vec{\bm{q}}_i \Vert\big) + \tanh\big(\phi_g\big(\bm{q}_{i} \oplus \Vert \bm{V}  \vec{\bm{q}}_i \Vert\big)\big) \bm{q}_{i}
\end{equation}

\begin{equation}
    \vec{\bm{x}}_i^{(t+1)} = \phi_{h}\big(\bm{q}_{i} \oplus \Vert \bm{V}  \vec{\bm{q}}_i \Vert\big) \circ \bm{U}\vec{\bm{q}}_i
\end{equation}
where $\oplus$ denotes concatenation, \(\phi_{s}, \phi_{g}, \phi_{h}: \mathbb{R}^{2F} \rightarrow \mathbb{R}^{F} \) are MLPs, and \(\bm{U}, \bm{V} \in \mathbb{R}^{F \times F}\) are trainable matrices.

\subsection{Defect-aware coordinate updating}
The defect-aware coordinate updating step aims to refine the atomic coordinates using two modules, RPV2Disp and Vec2Disp, which represent two distinct contributions to the coordinate update. Specifically, RPV2Disp converts the relative position vector $\vec{\bm{r}}_{ji}^{(t)}$ into a displacement, while Vec2Disp translates the vector representation $\vec{\bm{x}}_i^{(t+1)}$ into a displacement. Together, these determine the displacement of each atom at the current stage, as described by the following equations:
\begin{equation}
    \vec{\bm{d}}_i^{(\text{RPV})} = \sum_{v_j \in \mathcal{N}(v_i)} \phi_{q} \Big( \phi_v (\bm{x}_{j}^{(t+1)}) \circ \lambda_v(\Vert \vec{\bm{r}}_{ji}^{(t)} \Vert) \circ \gamma_v \big(E(m_i) + E(m_j)\big) \Big) \circ \frac{\vec{\bm{r}}_{ji}^{(t)}}{\Vert \vec{\bm{r}}_{ji}^{(t)} \Vert}
\end{equation}
\begin{equation}
    \vec{\bm{d}}_i^{(\text{Vec})} = \bm{W}_{\text{Vec}}\vec{\bm{x}}_i^{(t+1)}
\end{equation}
Here, \(\phi_v:\mathbb{R}^F\!\to\!\mathbb{R}^F\) and \(\phi_q:\mathbb{R}^F\!\to\!\mathbb{R}\) are MLPs; \(\gamma_v\) is the pair‐wise defect gate that re‐weights messages according to the marker pair \((m_i,m_j)\); and \(\bm{W}_{\text{Vec}} \in \mathbb{R}^{1 \times F}\) integrates all the vectors within \(\vec{\bm{x}}_i^{(t+1)}\). Finally, the coordinates are updated as follows:
\begin{equation}
    \vec{\bm{r}}_i^{(t+1)} = \vec{\bm{r}}_i^{(t)} + \vec{\bm{d}}_i^{(\text{RPV})} + \vec{\bm{d}}_i^{(\text{Vec})}
\end{equation}
The initial coordinate \(\vec{\bm{r}}_i^{(0)}\) is set to the atom coordinate of the unrelaxed structure. \textcolor{black}{The updated coordinates $\vec{\bm{r}}_i^{(t+1)}$ are equivariant to both rotation and translation, with a formal proof provided in Supplementary Note 1.}

\subsection{Message distribution and aggregation}
To establish a more effective global communication channel across the entire graph, we implement a message distribution and aggregation scheme using global node technology \cite{yang2025efficient}. The message distribution process propagates the global scalar and vector at the current step to each node using the following equations:
\begin{equation}
    \bm{x}_i^{(t)} = \phi(\bm{x}_i^{(t-1)} \oplus \bm{x}_\mathcal{G}^{(t-1)}) + \bm{x}_i^{(t-1)}
\end{equation}
\begin{equation}
    \vec{\bm{x}}_i^{(t)} = \bm{W}(\vec{\bm{x}}_i^{(t-1)} + \vec{\bm{x}}_\mathcal{G}^{(t-1)}) + \vec{\bm{x}}_i^{(t-1)}
\end{equation}
where \(\phi: \mathbb{R}^{2F} \rightarrow \mathbb{R}^F\) is an MLP, and \( \bm{W} \in \mathbb{R}^{F \times F} \) is a trainable matrix.

The message aggregation step updates the global scalar and vector based on the node representations at the current step, as described by the following equations:
\begin{equation}
    \bm{x}_\mathcal{G}^{(t+1)} = \phi\left(\left(\frac{1}{\vert \mathcal{G} \vert}\sum_{v_i \in \mathcal{G}} \bm{x}_i^{(t)}\right) \oplus \bm{x}_\mathcal{G}^{(t)}\right) + \bm{x}_\mathcal{G}^{(t)}
\end{equation}
\begin{equation}
    \vec{\bm{x}}_\mathcal{G}^{(t+1)} = \bm{W}\left(\left(\frac{1}{\vert \mathcal{G} \vert}\sum_{v_i \in \mathcal{G}} \vec{\bm{x}}_i^{(t)}\right) + \vec{\bm{x}}_\mathcal{G}^{(t)}\right) + \vec{\bm{x}}_\mathcal{G}^{(t)}
\end{equation}

\textcolor{black}{It is important to note that this global communication pathway does not incorporate interatomic distances and thus does not model short-range interactions directly. Instead, such interactions—including those modified by defects—are explicitly captured by the localized, distance-aware, defect-sensitive message passing mechanism (defect-aware message passing) described in the previous section.}

\subsection{Non-linear vector activation}
Non-linearity is essential for enhancing the expressive power of neural networks. Here, we introduce non-linearity into vector representations while preserving equivariance. Specifically, we first aggregate the \(F\) vectors within a node to obtain a consensus vector for each node:
\begin{equation}
    \vec{\bm{x}}_i^{\mathcal{G}} = \bm{W}_{p}\vec{\bm{x}}_i
\end{equation}
where \(\bm{W}_{p} \in \mathbb{R}^{1 \times F}\) integrates all vectors within \(\vec{\bm{x}}_i\) to produce the consensus vector \(\vec{\bm{x}}_i^{\mathcal{G}} \in \mathbb{R}^3\), capturing the overarching trend across all vectors in the node. Next, each vector $\vec{\bm{x}}_i^j$ within $\vec{\bm{x}}_i$ is updated as follows:
\begin{equation}
    \vec{\bm{x}}_i^j = \left\{
    \begin{array}{ll}
        \bm{W}^j \vec{\bm{x}}_i, & \text{if } \langle \vec{\bm{x}}_i^{\mathcal{G}}, \bm{W}^j \vec{\bm{x}}_i \rangle > 0  \\
        \bm{W}^j\vec{\bm{x}}_i + \vec{\bm{x}}_i^{\mathcal{G}}, & \text{otherwise}
    \end{array}
    \right.
\end{equation}
Here, $\bm{W}^j \in \mathbb{R}^{1 \times F}$, and \(\langle \cdot, \cdot \rangle\) denotes the dot product. The idea is that if the vectors align with the consensus trend, as indicated by a dot product greater than zero, they are considered significant and retained without modification. Conversely, vectors that diverge from the consensus trend (dot product less than or equal to zero) are considered potentially noisy or weakly informative and are softly regularized by adding the consensus vector. This adjustment encourages alignment with the dominant directional trend. Every time the vectors have been updated, we apply a non-linear vector activation to them.

\subsection{Implementation details}
The DefiNet model is implemented using PyTorch, and experiments are conducted on an NVIDIA RTX A6000 with 48 GB of memory. The training objective is to minimize the mean absolute error (MAE) loss between the ML-relaxed and DFT-relaxed structures, defined as follows:
\begin{equation}
    \mathcal{L} = \frac{1}{N} \frac{1}{M} \sum\limits_{i=1}^M \left\lvert \vec{\bm{r}}_i^{(T)} - \tilde{\bm{r}}_i \right\rvert
\end{equation}
where \(N\) and \(M\) denote the sample size and the number of atoms in each sample, respectively. Here, \(T\) represents the total number of layers in the model, and \(\tilde{\bm{r}}_i\) is the DFT-relaxed atomic coordinate. We use the AdamW optimizer with a learning rate of 0.0001 to update the model's parameters. Additionally, a learning rate decay strategy is implemented, reducing the learning rate if there is no improvement in coordinate MAE for 5 consecutive epochs.

\subsection{DFT calculations}
\label{sec:dft_calc}
Our calculations are performed using DFT with the Perdew-Burke-Ernzerhof (PBE) exchange-correlation functional, as implemented in the Vienna Ab Initio Simulation Package (VASP) \cite{kresse1996efficient}. The interaction between valence electrons and ionic cores is treated using the projector augmented wave (PAW) method \cite{blochl1994projector}, with a plane-wave energy cutoff of 500 eV. Initial crystal structures were taken from the Materials Project database. Given the large supercells required for defect calculations, structural relaxations were carried out using a $\Gamma$-point only Monkhorst-Pack grid. To prevent interactions between neighboring layers, a vacuum space of at least 15 $\rm \AA$ was introduced. During structural relaxation, atomic positions were optimized until the forces on all atoms were below 0.01 $\rm eV/\AA$, with an energy tolerance of $10^{-6}$ eV. For defect structures with unpaired electrons, we used standard collinear spin-polarized calculations, initializing magnetic ions in a high-spin ferromagnetic state, with the possibility of relaxation to a low-spin state during the ionic and electronic relaxation processes.

\section*{Data Availability}
The data that support the findings of this study are available in \url{https://zenodo.org/records/14027373}.

\section*{Code Availability}
The source code for DefiNet is available at \url{https://github.com/Shen-Group/DefiNet}.

\section*{Acknowledgements}
This research was supported by the Natural Science Foundation of Guangdong Province (Grant No. 2025A1515011487) and by the Ministry of Education, Singapore, under its Research Center of Excellence award to the Institute for Functional Intelligent Materials (I-FIM, project No. EDUNC-33-18-279-V12). K.S.N. is grateful to the Royal Society (UK, grant number RSRP\textbackslash R\textbackslash190000) for support.

\section*{Author Contributions}
Lei Shen, Kostya S. Novoselov, Pengru Huang, and Ziduo Yang designed the research. Ziduo Yang conduct the experiment. Ziduo Yang, Xiaoqing Liu, Xiuying Zhang, Pengru Huang, and Lei Shen analyzed the data and results. Ziduo Yang and Lei Shen wrote the manuscript together. All authors reviewed and approved the final version of the manuscript.

\section*{Competing Interests}
The authors declare no competing interests.


\bibliography{sn-bibliography}


\end{document}